\documentclass[conference,letterpaper]{IEEEtran}

\usepackage[utf8]{inputenc} 
\usepackage[T1]{fontenc}
\usepackage{url}
\usepackage{ifthen}
\usepackage{cite}
\usepackage{amsfonts}
\usepackage{indentfirst}
\usepackage{amsmath,amsthm}
\usepackage{amssymb}
\usepackage{color}
\usepackage{graphicx}
\usepackage[para]{threeparttable}
\usepackage{cases}
\usepackage{flushend}

\DeclareMathOperator{\Rnk}{Rank}
\DeclareMathOperator{\Norm}{Norm}

\usepackage[linesnumbered,algoruled,boxed,lined]{algorithm2e}

\newcommand{\etal}{\textit{et al.}}

\newcommand{\Fm}{\mathbb F_{q^m}}
\newcommand{\Fn}{\mathbb F_{q^n}}
\newcommand{\Fq}{\mathbb F_{q}}

\newtheorem{Definition}{Definition}
\newtheorem{Proposition}{Proposition}
\newtheorem{Remark}{Remark}
\newtheorem{Lemma}{Lemma}

\usepackage{threeparttable}
\usepackage{multirow}

\begin{document}
\title{New Communication Models and Decoding of 
Maximum Rank Distance Codes } 


\author{%
  \IEEEauthorblockN{Wrya K. Kadir}
  \IEEEauthorblockA{Department of Informatics\\ University of Bergen, Norway\\
Email: wrya.kadir@uib.no}
}


\maketitle

\begin{abstract}

In this paper an interpolation-based decoding algorithm to decode Gabidulin codes transmitted through a new communication model is proposed. The algorithm is able to decode rank errors beyond half the minimum distance by one unit. Also the existing  decoding algorithms for generalized twisted Gabidulin codes and additive generalized twisted Gabidulin codes are improved. 
\end{abstract}


\section{Introduction}

Delsarte \cite{Delsarte:1978aa}, Gabidulin \cite{Gabidulin1985} and Roth \cite{roth1991maximum} independently introduced \textit{rank metric codes}. Those rank metric codes that achieve Singleton-like bound are called \textit{maximum rank distance (MRD) codes}. Gabidulin codes are the most well known family of MRD codes.  
Later this family was generalized by Kshevetskiy and Gabidulin  \cite{kshevetskiy2005new} to \textit{generalized Gabidulin (GG)} codes. These codes are linear over $\Fn$. Sheekey in \cite{Sheekey} defined \textit{twsited Gabidulin (TG)} codes and established a way to generalize GG codes to linear MRD codes over a base fields and then he was followed by Lunardon \etal \cite{LTZ}, Otal and {\"O}zbudak \cite{Otal2017}, Trombetti and Zhou \cite{TrombettiZhou2019} and Sheekey \cite{Sheekey2020newMRD} to define \textit{generalized twisted Gabidulin (GTG)} codes, \textit{additive generalized twisted (AGTG)} codes, \textit{Trombetti-Zhou (TZ)} codes and \textit{new MRD codes by Sheekey}, repcetively.  For more constructions of MRD codes, please refer to \cite{Sheey2019}.

Efficient decoding  is required for the wide range of applications of MRD codes in  storage system \cite{roth1991maximum}, network coding \cite{SilvaKschischangKoetter} and cryptography \cite{GPT}. There are plenty of algorithms that decode Gabidulin codes up to half the minimum distance \cite{Gabidulin1985,Richter,Loidreau:2006aa,Tovohery2018} and some which  decode Gabidulin codes beyond half the minimum distance by considering restricted communication models \cite{Gab-philipchuk-symmetric-rank-error-error2004,philipchuk-Gab-symmetric-error2006,GABIDULIN-phlipchuk--symmetric-matrices-2006,Renner-Bartz-Sven-random-decoding-Gab,jerko-sidorenko-zeh-space-symmetric21}. The previously proposed restricted models, can generate error vectors that hold some structure and they do not look random.

Randrianarisoa in \cite{Tovohery2018} gave an interpolation-based decoding algorithm for Gabidulin codes and also for GTG codes. This idea is used later in  \cite{Kadir-li20},\cite{Li2019}, \cite{Kadir-li-Zullo-isit21} and \cite{kadir-li-Zullo-bfa21} to decode AGTG \cite{Otal2017}, Non-additive partition MRD codes \cite{Otal:2018aa}, TZ codes \cite{TrombettiZhou2019} and Hermitain Rank metric codes \cite{schmidt2018hermitian}, respectively. 

In this paper we decode Gabidulin codes beyond half the minimum distance and also improve the decoding algorithms for GTG in \cite{Tovohery2018} and AGTG  codes in  \cite{Li-kadir-wcc19,Kadir-li20} by making some delicate restrictions on the communication model. In the previously defined restricted models, the error vectors hold some specific structures, for instance symmetric error vectors \cite{Gab-philipchuk-symmetric-rank-error-error2004}, space-symmetric error vectors \cite{jerko-sidorenko-zeh-space-symmetric21}, but the channels in our model generate error vectors without any specific structure. Moreover, we use low rate GTG and AGTG codes at the end of this paper to decode error vectors with rank $\leq k$ where $k$ is the dimension of the code. 



\section{Preliminaries} 

\begin{Definition}
Let $q$ be a power of prime $p$ and $\Fm$ be an extension of the finite field $\Fq$. A $q$-polynomial is a polynomial of the form $L(x)=a_0x+a_1x^q+\cdots+a_{k-1}x^{q^{k-1}}$ over $\Fm$. If $a_{k-1}\neq 0$, then we say that $L(x)$ has $q$-degree $k-1$. 
 The set of all linearized polynomials of the form  $L(x)$ is denoted by $\mathcal{L}_k(\Fm)$.
\end{Definition}
When $q$ is fixed or the context is clear, it is also customary to speak of a \textit{linearized polynomial} as it satisfies the linearity property: $L(c_1x+c_2y)=c_1L(x)+c_2L(y)$ for any $c_1,c_2 \in\Fq$ and any $x,y$ in an arbitrary extension of $\Fm$. Hence a linearized polynomial $L(x)\in \mathcal{L}_k(\Fm)$ defines an $\Fq$-linear transformation $L$ from $\Fm$ to itself.
The rank of a nonzero linearized polynomial $L(x)=\sum_{i=0}^{n}a_ix^{q^i}$ over $\Fm$ is given by $\mbox{Rank}(L)=n-\mbox{dim}_{\Fq}(\mbox{Ker}(L))$, where $\mbox{Ker}(L)$ is the kernel of $L(x)$.

\begin{Proposition}\label{prop-Dickson-tovo}
Let $L(x)=\sum_{i=0}^{n-1}a_ix^{q^i}$  over $\Fm$ be a linearized polynomial with rank $t$. Then its associated Dickson matrix 
\small
\begin{equation}\label{EqDicksonmatrix}
D=\begin{pmatrix}
a_{i-j({\,\rm mod }n)}^{q^i}
\end{pmatrix}_{n\times n}=\begin{pmatrix}
a_0& a_{n-1}^{q}& \cdots& a_1^{q^{n-1}}\\
a_1& a_0^{q}&\cdots& a_2^{q^{n-1}}\\
\vdots&\vdots&\ddots&\vdots\\
a_{n-1}& a_{n-2}^{q}&\cdots& a_{0}^{q^{n-1}}
\end{pmatrix},
\end{equation}
\normalsize
has rank $t$ over $\Fm$ \cite{Tovohery2018}. Moreover, any $t\times t$ submatrix formed by $t$ consecutive rows and $t$ consecutive columns in $D$ is non-singular \cite{MENICHETTI-86,dickson-book}. 
\end{Proposition} 

\section{Maximum rank distance (MRD) codes}

The rank of a vector $a=(a_1,\ldots,a_n)$ in $\Fm^n$, denoted as $\mbox{Rank}(a)$, is the number of its  linearly independent components, that is the dimension of 
the vector space spanned by $a_i$'s over $\Fq$. The rank distance between two vectors $a,b\in \Fm^n$ is defined as $d_R(a,b)=\mbox{Rank}(a-b)$.

\begin{Definition}
A subset $\mathcal{C}\subseteq \Fm^n$ with respect to the rank distance is called a rank metric code. When $\mathcal{C}$ contains at least two elements, the minimum rank distance of $\mathcal{C}$ is given by $d(\mathcal{C})=\displaystyle\min_{\substack{A,B\in \mathcal{C},~A\neq B}}\{{{d_R}}(A,B)\}$. Furthermore, it is called a \textit{maximum rank distance (MRD) code} if it attains the Singleton-like bound $|\mathcal{C}|\leq q^{\min \{m(n-d+1),n(m-d+1)\}}$. 
\end{Definition}
The most famous MRD codes are Gabidulin codes \cite{Gabidulin1985} which were further generalized in \cite{roth1996tensor,kshevetskiy2005new}. The generalized Gabidulin (GG) codes $\mathcal{GG}_{n,k}$ with length $n\leq m$ and dimension $k$ over $\Fm$ is defined by the evaluation of 
\begin{equation}\label{eq-GG codes}
\bigg\{\sum_{i=0}^{k-1}f_ix^{q^{si}}\; |f_i\in \Fm\bigg\},
\end{equation}
where $(s,{m})=1$, on linearly independent points $\alpha_0,\alpha_1,\ldots,\alpha_{n-1}$ in $\Fm$.
The choice of $\alpha_i$'s does not affect the rank property and it is customary to exhibit Gabidulin codes and its generalized families without the evaluation points as in  \eqref{eq-GG codes}. For consistency with the parameters of MRD codes in \cite{Sheekey,TrombettiZhou2019,Otal2017}, through what follows we always assume $n=m$.

For a linearized polynomial $L(x)=\sum_{i=0}^{k}l_ix^{q^i}$ over $\Fn$, it is clear that $\mbox{Rank}(L)\geq n-k$ if $l_{k}\neq 0$. Gow and Quinlan in \cite[Theorem 10]{GOW20091778} (see also \cite{Sheekey}) characterize a necessary condition for $L(x)$ to have rank $n-k$ as below, see \cite{CSAJBOK2019109lin-poly,MCGUIRE201968} for other necessary conditions. 
\begin{Lemma}\label{Sheekey-lemma} \cite{GOW20091778}
		Suppose a linearized polynomial $L(x)=l_0x+l_1x^{q}+\cdots+l_kx^{q^{k}}$, $l_k\neq 0$, in $\mathcal{L}_n({\mathbb{F}_{q^n}})$ has $q^k$ roots in $\mathbb{F}_{q^n}$. Then 
		$\Norm_{q^n/q}(l_k)= (-1)^{nk}\Norm_{q^n/q}(l_0),$
		where $\Norm_{q^n/q}(x)=x^{1+q+\cdots + q^{n-1}}$ is the norm function from $\mathbb{F}_{q^n}$ to $\mathbb{F}_q$.
\end{Lemma}
According to Lemma \ref{Sheekey-lemma}, a linearized polynomial $L(x)$ of $q$-degree $k$ has rank at least $n-k+1$ if the condition in Lemma \ref{Sheekey-lemma} is not met. 
Sheekey \cite{Sheekey} applied Lemma \ref{Sheekey-lemma} and constructed a new family of $\Fq$-linear MRD codes, known as \textit{twisted Gabidulin (TG) codes}, and the generalized TG codes are investigated in \cite{LTZ} as follows:
	\begin{equation}\label{Eq-TGG}
		\mathcal{H}_{k,s}(\epsilon, h) = \left\{ \sum_{i=0}^{k-1}f_ix^{q^{si}} + \epsilon f_0^{q^{h}}x^{q^{sk}} \;|\; f_i \in \mathbb{F}_{q^n} \right\},
		\end{equation}
		where $n,k,s,h$ are positive integers such that $k<n$ and $(s,n)=1$.  Here $\epsilon$ is a nonzero element in $\Fn$ satisfying $\Norm_{q^{sn}/q^s}(\epsilon)\neq (-1)^{nk}$.
Later Otal and {\"O}zbudak \cite{Otal2017} further generalized this family by manipulating some terms of linearized polynomials and constructed the following $\mathbb{F}_{q_0}$-linear MRD codes, known as \textit{additive generalized twisted Gabidulin (AGTG) codes} 

		\begin{equation}\label{Eq-AGTG}
		\mathcal{A}_{k,s,q_0}(\epsilon,h)= \left\{ \sum_{i=0}^{k-1}a_ix^{q^{si}} + \epsilon a_0^{q_0^{h}}x^{q^{sk}} \;|\; a_i \in \mathbb{F}_{q^n} \right\},
		\end{equation}
		where $q=q_0^u$ and nonzero $\epsilon$ in $\Fn$ satisfies $\Norm_{q_0^{snu}/q_0^s}(\epsilon)\neq (-1)^{nku}$.

For the rest of this paper, we use the notation $[i]:=q^{si}$ for $i=0, \ldots, n-1$ , where $\mbox{gcd}(s,n)=1$, for simplicity.

\section{New Communication Models}\label{sec-models}
In this section we define two new communication models. The models contain two authorized parties as sender and receiver. The sender encodes his/her message and then an error vector with rank $t$ is added to the encoded message. The receiver will be able to decode the error vector and recover the message. Each  models uses a different form of interpolation polynomial to generate its corresponding error vector.

\subsection{First Model}\label{sec-model-A}
In this modes, a linearized polynomial of the form

\begin{align}
    e_{\theta_1,\theta_2}(x)&=\sum_{i=0}^{n-1}z_ix^{[i]},\; z_i\in \mathbb{F}_{q^{n}},\label{Eq-error-poly}\\&z_0^{[n/2]}-z_0=\alpha_{\theta_1},\label{Eq-error-poly-relation1}\\&z_{k-1}^{[n/2]}-z_{k-1}=\alpha_{\theta_2}\label{Eq-error-poly-relation2},
\end{align}

is used as the error interpolation polynomial  where $\theta_1,\theta_2\in [0,n-1]$ are the models'  public parameters.
We denote this model by $\mathcal{Q}_{\theta_1,\theta_2}$.
	
\subsection{Second Model}\label{sec-model-B}
In this model we have two cases:

\begin{itemize}
    \item \textbf{case 1.} Suppose  $n$ is an odd integer, then  \begin{equation}\label{Eq-symmetric-odd}
    b(x)=b_0x^{[0]}+\sum_{i=1}^{\frac{n-1}{2}}(b_ix^{[i]}+(b_ix)^{[n-i]}),
\end{equation}
is the error interpolation polynomial where  $\Tilde{b}=(b_0,\ldots,b_{n-1})$, $b_i\in\Fn$ and
\begin{equation}\label{Eq-symmetric-relation1}
    b_{n-i}=b_i^{[n-i]} \mbox{ for }  i=1,\ldots, \dfrac{n-1}{2}. 
\end{equation}

\item \textbf{case 2.} Suppose  $n$ is an even integer, then  \begin{equation}\label{Eq-symmetric-even}
    h(x)=h_0x^{[0]}+\sum_{i=1}^{\frac{n}{2}-1}(h_ix^{[i]}+(h_ix)^{[n-i-1]})+h_{n-1}x^{[n-1]},
\end{equation}

is the error interpolation polynomial where
    $\tilde{h}=(h_0,\ldots,h_{n-1})$, $h_i\in\Fn$,  and \begin{equation}\label{Eq-symmetric-relation2}
    h_{n-i-1}=h_i^{[n-i-1]} \mbox{for }  i=1,\ldots, \dfrac{n}{2}-1.
\end{equation}
\end{itemize}
Suppose $s(x)$ be one of the polynomials $e_{\theta_1,\theta_2},b(x)$ or $h(x)$. We use $s(x)$  such that	\begin{equation}\label{EqInterpolation--1}
s(\alpha_i)=e_i, \quad i=0, \ldots, n-1,
	\end{equation}
	where $e=(e_0,\ldots, e_{n-1})$ is the error vector and $\alpha_0,\ldots,\alpha_{n-1}$ are ordered linearly  independent points in $\mathbb{F}_{q^{n}}$ over $\Fq$.

\section{Decoding Gabidulin codes beyond half the minimum distance}

\vspace{-6pt}
\subsection{Encoding}\label{Subsec3.1}

	Let $\mathcal{GG}_{n,k}$, where $n$ is even and $k$ is odd, be a Gabidulin code with ordered $\Fq$-linearly independent evaluation points $\alpha_0,\alpha_1, \ldots, \alpha_{n-1}$. The encoding of
	a message $m=(m_0,\ldots, m_{k-1})$ is the evaluation of the following linearized polynomial at points $\alpha_0,\alpha_1, \ldots, \alpha_{n-1}$:
	\begin{equation}\label{eq-TZ-linearized-form}
	    f(x)=\sum_{i=0}^{k-1}m_ix^{[i]},
	\end{equation}
	
	Let 
	$\tilde{m}=(m_0,m_1,\ldots, m_{k-1}, 0, \ldots, 0)$ be a vector of length $n$ over $\mathbb{F}_{q^{n}}$ and $	M=
	\begin{pmatrix}
		\alpha_i^{[j]}
	\end{pmatrix}_{n\times n}$
	be the  \textit{Moore matrix} generated by $\alpha_i$'s, where $1\leq i, j\leq n-1$.
	Then the encoding of the message $m$ can be expressed as 
	\small
	\begin{equation}\label{EqTZ-Encoding}
	(m_0,m_1,\ldots, m_{k-1})\mapsto c=(f(\alpha_0), \ldots, f(\alpha_{n-1}))=\tilde{m}\cdot M^T,
	\end{equation}
	\normalsize
	where $M^T$ is the transpose of matrix $M$. In this process since only the first $k$ components of $\tilde{m}$ are nonzero, so only the first $k$ rows of $M$ are involved.

	\vspace{-6pt}
	\subsection{Decoding errors with rank $t\leq \frac{n-k+1}{2}$}\label{Subsec3.2}
	


	
	Let the error vector $e=(e_0,\dots,e_{n-1})$ of rank $t$ be added to the codeword $c=(c_0\ldots,c_{n-1})$ during transmission and let $r=(r_0\ldots,r_{n-1})=c+e$ be the received vector. 

	
		\smallskip
	
	
	\smallskip
	
	Suppose we use the communication model $\mathcal{Q}_{\theta_1,\theta_2}$ and let $e_{\theta_1,\theta_2}$ in \eqref{Eq-error-poly} be the error interpolation polynomial such that 
	 	\begin{equation}\label{EqInterpolation--2}
	e_{\theta_1,\theta_2}(\alpha_i)=e_i=r_i-c_i, \quad i=0, \ldots, n-1,
	\end{equation}
	where $\alpha_0,\ldots,\alpha_{n-1}$ are ordered linearly  independent points over $\Fq$  in $\mathbb{F}_{q^{n}}$.
	One can see that the error vector  $e$ is uniquely determined by the polynomial $e_{\theta_1,\theta_2}(x)$ and denote $z=(z_0,\ldots, z_{n-1})$. From
	\eqref{EqTZ-Encoding} and \eqref{EqInterpolation--2} it follows that
	$$
	r = c+e = (\tilde{m}+z)\cdot M^T.
	$$
	Since $M$ is nonsingular, this can be rewritten as 
	\begin{align*}
	  r \cdot (M^T)^{-1}=& (c_0, c_1, \,\ldots, c_{k-1}, 0, \ldots, 0) + \\
	  & (z_0,z_1, \ldots, z_{k-1}, z_k, \ldots, z_{n-1}).
		\end{align*}
	Let $\tilde{r}=(\eta_0, \ldots, \eta_{n-1})=r \cdot (M^T)^{-1}$, then the known coefficients $z_i$'s are
	\begin{equation} \label{EqInterpolation-3}
	(z_{k}, \ldots, z_{n-1}) = (\eta_{k}, \ldots, \eta_{n-1}),
	\end{equation}
	and we also have the auxiliary equations \eqref{Eq-error-poly-relation1} and \eqref{Eq-error-poly-relation2} which we will use later.

	\subsection{Reconstructing the interpolation polynomial $e_{\theta_1,\theta_2}(x)$}\label{SubSec3.3}
	
	
	
	\smallskip
	

	Let
	\begin{equation}\label{Eq-DM-G-Simplified}
	E=\begin{pmatrix}z^{[j]}_{i-j~({\rm mod~}n)} 
	\end{pmatrix}_{n\times n}
	= \left(E_0 \,\, E_1 \,\, \ldots \,\, E_{n-1}\right),
	\end{equation} 
	be the Dickson matrix associated with the linearized polynomial $e_{\theta_1,\theta_2}(x)$, 
	where the indices $i, \,j$ run through $\{0, 1, \ldots, n-1\}$ and 
	$E_{j}$ is the $j$-th column of $E$.
	
	
	According to Proposition \ref{prop-Dickson-tovo}, since $e_{\theta_1,\theta_2}(x)$ has rank $t$, so $E$ has rank $t$ and any $t\times t$ sub-miatrix of $E$ which contains $t$ consecutive rows and columns is nonsingular. Hence the first column $E_0$ can be written as the linear combination of columns $E_1\ldots, E_{t}$ as	
$
E_{0} = \gamma_1 E_1 +\gamma_2 E_2 + \cdots + \gamma_{t} E_{t},
$
where $\gamma_1,\ldots, \gamma_{t}$ are elements in $\mathbb{F}_{q^{n}}$.
Then we can obtain the following recursive equations
\begin{equation}\label{Eq-Gsub}
z_{i} = \gamma_1 z^{[1]}_{i-1} + \gamma_2 z^{[2]}_{i-2} + \cdots + \gamma_t z^{[t]}_{i-t}, \quad 0\leq i < n.
\end{equation} 
Due to the relation in \eqref{EqInterpolation-3}, we already know  $z_{k},\ldots, z_{n-1}$. These known coefficients leads us to the following linear recursive equation 

\begin{equation}\label{Eq-Gsub-4}
z_{i} = \gamma_1 z^{[1]}_{i-1} + \gamma_2 z^{[2]}_{i-2} + \cdots + \gamma_t z^{[t]}_{i-t}, \,\, k+t\leq  i <n,
\end{equation} where $\gamma_0\ldots, \gamma_{t}$ are unknowns. 
	 In \cite{Sidorenk}, the $q$-linearized shift register is given and the above recursive relation \eqref{Eq-Gsub-4} can be seen as its generalized version.  
	Here $(\gamma_1, \ldots, \gamma_{t})$ is the connection vector of the shift register. 
We call the equation \eqref{Eq-Gsub-4} as the \textit{key equation} for the decoding algorithm in this paper and due to the properties of shift register, finding $\gamma_1,\ldots,\gamma_t$ leads us to find the unknown coefficients $z_0\ldots, z_{k-1}$,  recursively. The most complex task in our decoding algorithm is finding $\gamma_1,\ldots,\gamma_t$ and then the remaining task (calculating unknown $z_i$'s) will be a recursive process. 
	We consider $\Rnk(e)=t \leq  \frac{n-k+1}{2}$, i.e., $2t+k\leq n+1$, and the task of finding $\gamma_1\ldots,\gamma_t$ via \eqref{Eq-Gsub-4} is divided into two cases:
	
	\noindent\textit{Case 1:} If $2t+k< n+1$. In this case, \eqref{Eq-Gsub-4} contains 
	$n-k-t\geq t$ affine equations 
	and $t$ variables $\gamma_1, \ldots, \gamma_{t}$, which has rank $t$. Hence the variables $\gamma_1, \ldots, \gamma_{t}$ can be uniquely determined. Here any Gabidulin decoder can be applied,  but here we assume the code has high code rate, for which the Berlekamp-Massey algorithm is more efficient and it has polynomial time complexity. 

	\smallskip
	
	\noindent\textit{Case 2:} If $2t+k=n+1$. In this case \eqref{Eq-Gsub-4} is an under-determined system of 
	 $n-k-t=t-1$ equations with $t$ variables $\gamma_1, \ldots, \gamma_{t}$. A set of solutions $(\gamma_1,\ldots,\gamma_t)$  with dimension one can be expressed of the form
	 
	\begin{equation}\label{Eq-gamma-BM-dim1}
	    \gamma+X \gamma'
	=(\gamma_1+X\gamma'_1, \ldots, \gamma_{t}+X \gamma'_{t}),
	\end{equation}
	
	 where $\gamma, \gamma'$ are fixed elements in $\mathbb{F}_{q^{n}}^t$ and $X$ runs through $\mathbb{F}_{q^{n}}$.
	The modified BM algorithm in \cite[Th. 10]{Sidorenk} can give the solution with a free variable $X$.
	
	
	\smallskip
	
If we take $i=0$ and $i=k+t-1$ in \eqref{Eq-Gsub-4} and substitute the solution  \eqref{Eq-gamma-BM-dim1}, then we get 
	\begin{equation}\label{eq20}
	z_0  =  \delta_0 + \delta_1 X ,
	\end{equation}
	and \begin{equation}\label{eq22}
	 	z_{k+t-1} =  \delta_2+\delta_3 X+ (\gamma_{t}+\gamma'_{t}X)z_{k-1}^{[t]}, 
	\end{equation}
	where in \eqref{eq20} and \eqref{eq22}, $z_0,z_{k-1}$ and $X$ are the only unknowns and $\delta_0,\delta_1,\delta_2,\delta_3$ are derived from $\gamma, \gamma'$ and known coefficients $z_k,\ldots,z_{n-1}$. $X=-\gamma_t/\gamma_t$ if $\gamma_{t}+\gamma'_{t}X=0$ and this solution can be verified by $\delta_2,\delta_3$ and a known coefficient $z_i$ in \eqref{eq22}.   Substituting \eqref{eq20} in \eqref{Eq-error-poly-relation1} gives \begin{equation}\label{eq23}
 	    \tau_0 X^{[n/2]} +\tau_1 X +\tau_2=0. 
 	\end{equation}
 	As the next step, we rise both sides of \eqref{eq22} to the  $[-t]$-th power and obtain 
 	
 	\begin{equation}\label{eq24}
 	   z_{k-1}=\dfrac{a_1+a_2X^{[-t]}}{a_3+a_4X^{[-t]}}.
 	\end{equation}
 	We also substitute \eqref{eq24} in \eqref{Eq-error-poly-relation2} and rise both sides to the $[t]$-th power to get 
 	\begin{equation}\label{eq25}
 	u_1X^{[n/2]+1}+u_2X^{[n/2]}+u_3X+u_4=0. 
 	\end{equation}
 	Finally, one can substitute \eqref{eq23} into \eqref{eq25} and obtain the following quadratic polynomial equation over $\mathbb{F}_{q^{n}}$
 	\begin{equation}\label{eq26}
 	  \mu_1 X^2+\mu_2 X+\mu_3=0.
 	\end{equation}
 	If $\mu_1=0$, then $X=-\mu_3/\mu_2$ and if $\mu_1\neq 0$, equation \eqref{eq26} can be reduced to 
 	\begin{equation}\label{eq27}
 	    X^2+rX+s=0,
 	\end{equation} where $r=\mu_2/\mu_1$ and $s=\mu_3/\mu_1$. When the characteristic of $\mathbb{F}_q$ is odd,  equation \eqref{eq27} can be solved explicitly as follows: 
\begin{itemize}
    \item[a)] if ${r^2-4s}$ is a quadratic residue in  $\mathbb{F}_{q^{n}}$, then it has two solutions $X=\frac{-r\pm\sqrt{r^2-4s}}{2}$;
    \item[b)] if $r^2=4s$, then it has a single solution $X=-r/2$;
    \item[c)] it has no solution in      $\mathbb{F}_{q^{n}}$  otherwise.
\end{itemize}
When the characteristic of $\Fq$ is two, we have the following cases:
\begin{enumerate}
    \item if $r=0$, it has a single solution $X=s^{2^{nl-1}}$,  where $q=2^l$;
    \item if $r\neq 0$, the equation \eqref{eq27} can be reduced to $y^2+y=\beta$, where $X=ry$ and $\beta=s/r^2$. Then $y^2+y=\beta$ has 
    \begin{itemize}
        \item no zero if $\sum_{i=0}^{n-1}\beta^{2^{i}}=1$;
        \item two zeros of the form $W=\sum_{j=1}^{n-1} \beta^{2^{j}}(\sum_{k=0}^{j-1}c^{2^{k}})$ and $W+1$ where $\sum_{i=0}^{n-1}\beta^{2^{i}}=0$ and $c$ is any fixed element such that $\sum_{i=0}^{n-1}c^{2^{i}}=1$.
    \end{itemize}
\end{enumerate}

	
%
	
	\smallskip

	We expect our quadratic equation
	to have roots $X$ in $\mathbb{F}_{q^{n}}$ that lead to solutions $\gamma+X \gamma'$ in \eqref{Eq-Gsub-4} and $z_0$ in \eqref{eq20}. With the coefficients $\gamma_1, \ldots, \gamma_{t}$ 
	and also the initial state $z_{n-1}, \ldots, z_{n-t}$, 
	one can recursively compute
	$z_1, \ldots, z_{k-1}$  according to \eqref{Eq-Gsub}.
	Note that even if the equation \eqref{eq26} has two different solutions, 
	they don't necessarily lead to correct coefficients of the error interpolation polynomial.
	In fact, by the expression of the Dickson matrix of $e_{\theta_1,\theta_2}(x)$, the correct 
	$e_{\theta_1,\theta_2}(x)$ should have the sequence $(z_{n-1}, \ldots, z_{n-t}, \ldots )$ with period $n$.
	In other words, if the output sequence has period $n$, we know that
	the corresponding polynomial $e_{\theta_1,\theta_2}(x)$ is the desired error interpolation polynomial.

\section{An improvement of the decoding of GTG and AGTG codes}
In the interpolation-based decodings of GTG and AGTG codes in \cite{rosenthal2017decoding,Tovohery2018,Li-kadir-wcc19} and\cite{Kadir-li20}, when the rank of the error vector $e$ is $t< \frac{n-k}{2}$, one can use any decoder of a Gabidulin code $\mathcal{GG}_{n,k+1}$ to recover the message. But when $t=\frac{n-k}{2}$, the problem of decoding the error vector is transformed to the problem of solving the projective polynomial $P(x)=x^{q^w+1}+u_1x+u_2=0$ over $\Fn$.  In the following, we show that how one can decode GTG and AGTG codes more efficiently if he/she  communicates via the communication model  $\mathcal{Q}_{\theta_1,\theta_2}$.  Moreover, we show that one will be able to decode any error vector with  rank $t\leq k$ added to a low rate GTG and AGTG code if one uses the second communication model. In this paper by a low rate code we mean a code with $k\leq \lceil\frac{n-1}{2}\rceil$.

\subsection{Decoding GTG and AGTG codes}\label{subsection-GTG-AGTG}

Here we explain an improvement of the decoding algorithm for GTG codes and the same procedure can be applied to AGTG codes with some minor differences. In this subsection we assume $n$ as an even positive integer.   To be self-contained, we recall the decoding algorithm from \cite{Kadir-li20} where the general communication model is replaced by the communication model $\mathcal{Q}_{\theta_1,\theta_2}$. 

\subsubsection{Encoding}
The encoding of
	a message $m=(m_0,\ldots, m_{k-1})$ is the evaluation of the following linearized polynomial at ordered points $\alpha_0,\alpha_1, \ldots, \alpha_{n-1}$:
	\begin{equation}\label{eq-GTG-linearized-form}
	   f(x)=\sum_{i=0}^{k-1}m_ix^{[i]}+\epsilon m_0^{q^h}x^{[k]}.
	\end{equation}
	
	Then the encoding of GTG codes can be expressed as 
	\small
	\begin{equation}\label{EqTZ-Encoding-1}
	(m_0,m_1,\ldots, m_{k-1})\mapsto c=(f(\alpha_0), \ldots, f(\alpha_{n-1}))=\tilde{m}\cdot M^T,
	\end{equation}
	\normalsize
	where 	$\tilde{m}=(m_0,\ldots, m_{k-1}, \epsilon m_0^{q^h}, 0, \ldots, 0)$.

\subsubsection{Decoding}
	Let the error vector $e=(e_0,\dots,e_{n-1})$ of rank $t$ be added to the codeword $c=(c_0\ldots,c_{n-1})$ during transmission and let $r=(r_0\ldots,r_{n-1})=c+e$ be the received vector. 
	Take $e(x)$ be the error interpolation polynomial  of the form given in \eqref{Eq-error-poly} where instead of \eqref{Eq-error-poly-relation2} we have
	\begin{equation}\label{Eq-error-poly-relation3}
	  z_{k}^{[n/2]}-z_{k}=\alpha_{\theta_2}.
	\end{equation}
	Then 
	\begin{equation}\label{EqInterpolation-1}
	e(\alpha_i)=e_i=r_i-c_i, \quad i=0, \ldots, n-1.
	\end{equation}
	As we mentioned before, $e$ is uniquely determined by the polynomial $e(x)$ and denote $z=(z_0,\ldots, z_{n-1})$. From
	\eqref{EqTZ-Encoding} and \eqref{EqInterpolation--2} it follows that
	$$
	r = c+e = (\tilde{m}+z)\cdot M^T.
	$$

	This is equivalent to
	\begin{align*}
	  r \cdot (M^T)^{-1}=& (m_0, m_1, \,\ldots, m_{k-1},\epsilon m_0^{q^h}, 0, \ldots, 0) + \\
	  & (z_0,z_1, \ldots, z_{k-1}, z_k,z_{k+1}, \ldots, z_{n-1}).
		\end{align*}
	Letting $\tilde{r}=(\eta_0, \ldots, \eta_{n-1})=r \cdot (M^T)^{-1}$, we obtain 
	\begin{equation} \label{EqInterpolation-3-1}
	(z_{k+1}, \ldots, z_{n-1}) = (\eta_{k+1}, \ldots, \eta_{n-1}),
	\end{equation}
	and we also have the relations \eqref{Eq-error-poly-relation1} and \eqref{Eq-error-poly-relation3}. In \eqref{EqInterpolation-3-1} we have $n-k-1$ known coefficients $z_i$'s, while in \eqref{EqInterpolation-3} we had $n-k$ known coefficients $`_i$'s.
\subsubsection{Reconstructing the interpolation polynomial $e(x)$}

If we write the  $0$th column $E_0$ of the Dickson matrix associated to $e(x)$ as the linear combination of $E_1,\ldots,E_t$ we will get the recursive equation 
\begin{equation}\label{Eq-Gsub-1}
z_{i} = \gamma_1 z^{[1]}_{i-1} + \gamma_2 z^{[2]}_{i-2} + \cdots + \gamma_t z^{[t]}_{i-t}, \quad 0\leq i < n,
\end{equation} same as \eqref{Eq-Gsub}, where the subscripts in $z_i$'s are taken modulo $n$.
Recall that the elements $z_{k+1},\ldots, z_{n-1}$ are known from \eqref{EqInterpolation-3-1}. 
Hence we obtain the following linear equations to replace the key equation in \eqref{Eq-Gsub-4}, with known coefficients $z_i$ and  variables $\gamma_1, \ldots, \gamma_{t}$:
\begin{equation}\label{Eq-Gsub-4-1}
z_{i} = \gamma_1 z^{[1]}_{i-1} + \gamma_2 z^{[2]}_{i-2} + \cdots + \gamma_t z^{[t]}_{i-t}, \,\, k+t+1\leq  i <n.
\end{equation}

	For an error vector with $\Rnk(e)=t \leq  \frac{n-k}{2}$, i.e., $2t+k\leq n$, we can divide the discussion into two cases.
	
	\noindent\textit{Case 1:} $2t+k< n$. In this case, \eqref{Eq-Gsub-4-1} contains 
	$n-k-t-1\geq t$ affine equations 
	in variables $\gamma_1, \ldots, \gamma_{t}$, which has rank $t$. Hence the variables $\gamma_1, \ldots, \gamma_{t}$ can be uniquely determined. Any Gabidulin $\mathcal{GG}_{n,k+1}$ decoder can be applied.  Here we assume the code has high code rate, for which the Berlekamp-Massey algorithm gives a better complexity. 
	Although the recurrence equation \eqref{Eq-Gsub-4-1} is a generalized version of the ones in \cite{Richter} and \cite{Sidorenk}, the modified Berlekamp-Massey algorithm can be applied here to recover the 
	coefficients $\gamma_1, \ldots, \gamma_{t}$.

	\smallskip
	
	\noindent\textit{Case 2:} $2t+k=n$. In this case \eqref{Eq-Gsub-4-1} 
	gives $n-k-t-1=t-1$ independent affine equations in variables $\gamma_1, \ldots, \gamma_{t}$.
	For such an under-determined system of linear equations, we will have a set of 
	solutions $(\gamma_1, \ldots, \gamma_{t})$
	that has dimension $1$ over $\mathbb{F}_{q^{n}}$. 
	Namely, the solutions will be  of the form 
	$$
	\gamma+X \gamma'
	=(\gamma_1+X\gamma'_1, \ldots, \gamma_{t}+X \gamma'_{t}),
	$$ where $\gamma, \gamma'$ are fixed elements in $\mathbb{F}_{q^{n}}^t$ and $X$ runs through $\mathbb{F}_{q^{n}}$.
	As shown in \cite[Th. 10]{Sidorenk}, the solution can be derived from the modified BM algorithm with a free variable $X$.
	
	\smallskip
	
	Observe that in \eqref{Eq-Gsub-1}, by taking $i=0$ and $i=k+t$ and substituting the solution $\gamma+X\gamma'$, one gets the following two equations
	\begin{equation}\label{eq20-1}
	z_0  =  \delta_0' + \delta_1' X ,
	\end{equation}
	and \begin{equation}\label{eq22-1}
	   	z_{k+t} =  \delta_2+\delta_3 X+ (\gamma_{t}+\gamma'_{t}X)z_{k}^{[t]},
	\end{equation}
	where in \eqref{eq20-1} and \eqref{eq22-1}, $z_0,z_k$ and $X$ are unknowns. Using equations \eqref{Eq-error-poly-relation1},\eqref{Eq-error-poly-relation3}, \eqref{eq20-1} and \eqref{eq22-1} instead of \eqref{Eq-error-poly-relation1},\eqref{Eq-error-poly-relation2}, \eqref{eq20} and \eqref{eq22} and going through  the same procedure in Subsection \ref{SubSec3.3}, we can get a quadratic equation of the form \begin{equation}\label{eq26-1}
 	  \mu_1 X^2+\mu_2 X+\mu_3=0.
 	\end{equation}
	which can be solved in polynomial time as discussed in Subsection \ref{SubSec3.3}. Hence, if the communication parties use the model $\mathcal{Q}_{\theta_1,\theta_2}$  to transfer their messages, then  GTG and AGTG codes can be decoded with less time complexity. 

	
\section{Decoding error rank vectors with  any rank $t\leq k$ }\label{Subsection-6}
In this subsection we consider the second communication model described in \ref{sec-model-B} , but the generated error vectors are still look random and they can have any rank up to $n$. 




In the decoding of GTG codes in  Subsection \ref{subsection-GTG-AGTG}, let $\tilde{r}=(\eta_0, \ldots, \eta_{n-1})=r \cdot (M^T)^{-1}$, then we obtain 
\begin{equation}\label{EqInterpolation-3-1-last}
	(z_{k+1}, \ldots, z_{n-1}) = (\eta_{k+1}, \ldots, \eta_{n-1}),
	\end{equation}
	and also based on the definition of GTG codes we have an auxiliary equation 
	\begin{equation}\label{Eq-AUX-g0-gk}
	-\epsilon z_0^{q^h} + z_k = \eta_k-\epsilon \eta_0^{q^h},
	\end{equation}
	since $\epsilon m_0^{q^h}+z_k=\eta_k,$ and $m_0+z_0=\eta_0$. Let $k\leq \lceil \frac{n-1}{2}\rceil$. If we use \eqref{Eq-symmetric-odd}  (\eqref{Eq-symmetric-even}) as the error interpolation polynomial, one can  employ \eqref{Eq-symmetric-relation1} (\eqref{Eq-symmetric-relation2})   and directly obtain  $z_1,\ldots, z_k$ from the known coefficients in  \eqref{EqInterpolation-3-1-last}. The only remaining unknown coefficient $z_0$ can be calculated using the auxiliary equation \eqref{Eq-AUX-g0-gk} since $z_k$ is already calculated.

Hence, by restricting the error interpolation polynomial we can decode any rank error vector with rank $t\leq k$  added to a low rate GTG (AGTG) code.

\begin{Remark}
In \cite{jerko-sidorenko-zeh-space-symmetric21}, an application of space-symmetric rank errors in code-based cryptography is proposed. But space-symmetric rank errors similar to symmetric rank errors \cite{Gab-philipchuk-symmetric-rank-error-error2004}, contain some structures and this may lead to a new structural attack.  If we use rank error vectors  defined in Subsection \ref{Subsection-6} instead of space-symmetric rank errors and use GTG codes instead of Gabidulin codes in GPT variants \cite{loidreau2016evolution} and \cite{loidreau2017new}, we can avoid potential structural attacks and  possibly get the same key size found in \cite[Section VI.]{jerko-sidorenko-zeh-space-symmetric21}. This will be investigated in future works. 
\end{Remark}

\begin{Remark}
The advantage of the model $\mathcal{Q}_{\theta_1,\theta_2}$ or even the second model \ref{sec-model-B} is that it can generate  error vectors that do not carry  a specific structure since the structured coefficients' vector of the error interpolation polynomial goes through an interpolation process on linearly independent points. Even in subsection VI. the error space has dimension $n/2$ but it contains error with high or low ranks with no specific structure. 
So based on this observation, to find more suitable rank-based scheme,  besides looking for new MRD codes and find the most efficient one, one can also look for new communication models with higher error correctability.   
\end{Remark}

\section{Conclusion}

In this paper we made some delicate restrictions on  the communication model and decode Gabidulin codes beyond half the minimum distance by one unit in polynomial time. The error vectors which are added to the codewords in our model, do not carry a specific structure. Moreover, we improved the decoding algorithms for GTG and AGTG codes proposed in \cite{Tovohery2018} and  \cite{Kadir-li20}, if two parties communicate through the first defined  models. We are also able to decode any error vector with any rank $t\leq k$ added to low rate ($k\leq \lceil\frac{n-1}{2}\rceil $) GTG and AGTG codes if we employ the second communication model.

\section*{Acknowledgment}
The author would like to thank Dr. Chunlei Li and  Dr. Ferdinando Zullo for their helpful advice  and also the anonymous reviewers for their valuable suggestions and comments.


\bibliographystyle{IEEEtran}
\bibliography{IEEEabrv,RankMetricCodes}

\end{document}